\documentclass[5p, preprint]{elsarticle}
\journal{Journal of Systems and Software}

\usepackage{amsmath,amssymb,amsfonts}
\usepackage{algorithmic}
\usepackage{graphicx}
\usepackage{textcomp}
\usepackage{xcolor}
\usepackage{graphicx}
\usepackage{balance}
\usepackage{dblfloatfix}
\usepackage[utf8]{inputenc}

\usepackage{graphicx}
\usepackage{tikz}
\usepackage{float}
\usepackage{pifont}

\usepackage{paralist}

\newcommand{\xmark}{\ding{55}}%
\newcommand{\bi}{\begin{compactitem}}
\newcommand{\ei}{\end{compactitem}}

\usepackage{color, colortbl}
\newcommand{\be}{\begin{enumerate}}
\newcommand{\ee}{\end{enumerate}}

\newcommand{\fig}[1]{Figure~\ref{fig:#1}}

\usepackage{pgfplots}
\usetikzlibrary{patterns}
\usetikzlibrary{arrows}
\usetikzlibrary {positioning}
\tikzset{main node/.style={circle,fill=white!20,draw,minimum size=0.5cm,inner sep=0pt}}
\usepackage[tikz]{bclogo}
{\noindent\begin{minipage}[c]{\linewidth}%
\begin{bclogo}[couleur=gray!20,%
                arrondi=0.1,logo=\bctrombone,%
                ombre=true]{{\small ~#1}}}%
{\end{bclogo}\vspace{2mm}\end{minipage}}

\usepackage{enumitem}
\usepackage{subfigure}
\usepackage{xcolor}
\usepackage{hyperref}
\usepackage{dblfloatfix} 
\usepackage{tikz}
\usepackage{bchart}
\setlist[description]{leftmargin=1cm}
\usepackage{wrapfig}
\usepackage{multirow}

\definecolor{lightgray}{gray}{0.8}
\definecolor{LightCyan}{rgb}{0.88,1,1}
\definecolor{darkgray}{gray}{0.6}
\definecolor{Gray}{rgb}{0.88,1,1}
\definecolor{Gray}{gray}{0.85}
\definecolor{Blue}{RGB}{0,29,193}
\definecolor{MyDarkBlue}{rgb}{0,0.08,0.45} 
\definecolor{pink}{RGB}{231,95,110}
\definecolor{lightergray}{rgb}{0.85, 0.85, 0.85}
\definecolor{lightestgray}{rgb}{0.95, 0.95, 0.95}
\definecolor{ao(english)}{rgb}{0.0, 0.5, 0.0}

\newenvironment{result}
{\vspace{0.15cm}
\noindent\begin{minipage}{\linewidth}
\begin{center}
\arrayrulecolor{lightergray}
\begin{tabular}{|p{0.95\linewidth}|}
\hline%
\rowcolor{gray!30}%
\textbf{Result:}~%
}
{\\\hline
\end{tabular}
\end{center}
\end{minipage}
\vspace{0.15cm}
}
\newcommand{\quart}[4]{\begin{picture}(100,6)
{\color{black}\put(#2,3){\color{black}\circle*{4}}\put(#1,3){\line(1,0){#3}}}\end{picture}}

\def\BibTeX{{\rm B\kern-.05em{\sc i\kern-.025em b}\kern-.08em
    T\kern-.1667em\lower.7ex\hbox{E}\kern-.125emX}}

\begin{document}
\begin{frontmatter}
\title{Communication and Code Dependency Effects on Software Code Quality: \\ (An Empirical Analysis of Herbsleb Hypothesis)}

\author[ncsu]{Suvodeep Majumder}
\ead{smajumd3@ncsu.edu}
\author[ncsu]{Joymallya Chakraborty}
\ead{jchakra@ncsu.edu}
\author[ncsu]{Amritanshu Agrawal}
\ead{aagrawa8@ncsu.edu}
\author[ncsu]{Tim Menzies}
\ead{timm@ieee.org}
\address[ncsu]{Department of Computer Science, NC State University, NC}

\begin{abstract}


Prior literature has suggested that in many projects 80\% or more of the contributions are made by a small  called group of  around 20\% of
the development team.  Most prior studies deprecate a reliance on such a small inner group
of ``heroes'', arguing that it causes bottlenecks in development and communication. Despite this, such projects are very   common   in open source  projects.  So what exactly is the impact of ``heroes'' in code quality?

Herbsleb argues that if  code is strongly connected yet their developers are not, then that code will be buggy. To test the Hersleb hypothesis,   we develop  and apply two metrics of (a)~``social-ness' and (b)~``hero-ness'' that measure  (a)~how much one developer comments on the issues of another; and (b)~how much one developer changes another developer's code (and ``heroes'' are those that change the most code, all around the system). In a result endorsing the Hersleb hypothesis, in over 1000 open source projects, we find that ``social-ness'' is a statistically stronger indicate for code quality (number of bugs) than ``hero-ness''.   

Hence we say that debates over the merits of ``hero-ness'' is subtly misguided. Our results suggest that the real benefits of these so-called ``heroes'' is not so much the code they generate but the pattern of communication required when  the interaction between a large community of programmers passes through a small group of centralized developers. To say that another way, {\em to build better code, build better communication flows between core developers and the rest}.

In order to allow other researchers to confirm/improve/refute our results, all our scripts and data are available, on-line at  \href{https://github.com/Anonymous633671/A-Comparison-on-Communication-and-Code-Dependency-Effects-on-Software-Code-Quality}{our online Github repository}.

{\bf The authors declare that they have no known competing financial interests or personal relationships that could have appeared to influence the work reported in this paper.}

\end{abstract}

\end{frontmatter}

\section{Introduction}
\label{sec:intro}
 Developers talk amongst themselves to coordinate development activities such as prioritize and assign tasks, comprehend requirements better, discuss candidate solutions to complex problems, and like. Healthy communication is likely to result in good quality software.
Sound team communication is vital to the success of any software project involving multiple developers.
James Herbsleb~\cite{Herbsleb:2014}, in his ICSE’14 keynote said:
\begin{quote}
{\em If two code sections communicate but the programmers of those two sections do not, then that code section is more likely to be buggy.}\end{quote}

Many prior literature shows in many projects 80\% or more of the contributions are made by only 20\% of the developers. These developers are often referred to "hero" developers and the projects as ``hero'' projects. In the literature, hero projects are deprecated  since it is said, they are like bottlenecks that slow down the  project development process and causes information loss~\cite{bier2011online, boehm2006view, hislop2002integrating, morcovcomplex, wood2005multiview, fitzgerald2003making}. 
From the perspective
of Herbsleb hypothesis,
``hereos'' are a problem if they
result in less communication between developers.

Recent studies have motivated a re-examination of the implications of heroes. In 2018,  Agrawal et al.~\cite{Agrawal_2018} studied 661 open source projects and  171 in-house proprietary projects. In that sample, over 89\% of projects were hero-based\footnote{This text    use ``hero'' for   women and men since recent publications use it to denote admired people of all genders-- see bit.ly/2UhJCek.}. Since the widespread prevalence of heroes is  at odds with established wisdom in the SE literature.  The usual stance is to warn against heroes since they may become bottlenecks in development and communication~\cite{bier2011online, boehm2006view, hislop2002integrating, morcovcomplex, wood2005multiview, ricca2010heroes, robles2006contributor, capiluppi2007adapting}. Hence in software engineering research, it is a pressing issue to understand why so many projects are hero-based and what are implication of communication between these developers have on code quality (introduction of bugs in the code base). 
In most projects only a small group of developers are making the most contributions (writing the codes), then what happens to the code quality when they do and do not communicate with each other properly? Also, what about the code quality of developers who do not contribute much?

To that end, this paper tests the ``Herbsleb Hypothesis'':
\bi
    \item We collect data on
    over  1000 open source Github projects.
    \item We show that a majority of our projects are hero projects (i.e., majority percentage of contributions are made by only a small percentage of developers)
    \item We  explain {\em why} heroes are  important. As shown below, 
     when developer's codes interact with each other, if there is significant social interaction (social communication) between them as well they tends to introduce much less bugs into the code base than when there is not. 
\ei
As part of this research, we ask three research questions - 

{\bf RQ1:} {\em How common are projects with highly centralized social and code communication structure (i.e., hero projects)?} Here we see a overwhelming presence of  projects with highly centralized social and code communication structure in open source software community. That is, usually, there are very few people in each project responsible for most of the work.

{\bf RQ2:} {\em What impact  does high social and code communication have  on  code quality?} Reflecting on who writes most of the code is just as insightful as reflecting on who participates in most of the discussion about the code.

{\bf RQ3:} {\em Do the results support Herbsleb Hypothesis?} Our conclusion
will be ``yes''.

The rest of the paper is structured as follows. Section~\ref{sec:Background And Prior Work} provides background information that directly relates to our research questions, in addition to laying out the motivation behind our work. Section~\ref{sec:Data Collections} explains the data collection process and in  Section~\ref{sec:Experimental Setup}, a detailed description of our experimental setup and data is given, along with our performance criteria for evaluation is presented. It is followed by Section~\ref{sec:Results} the results of the experiments and answers to some research questions. threats to validity. Threats to validity are discussed in Section~\ref{sec:Threats to Validity}, which is followed by our conclusions, in   Section~\ref{sec:Conclusion}.

\subsection{Relation to Prior Work}
This paper extends prior work by  Agrawal et al.~\cite{Agrawal_2018}. That prior worked only looked at half the data explored here. Also, that work failed to detect the importance of the social effects reported here (so they could show empirically that hero projects existed but they did not recognize the important underlying social effects).

\begin{figure}[!b]
    \begin{center}
    \includegraphics[width=3.5in]{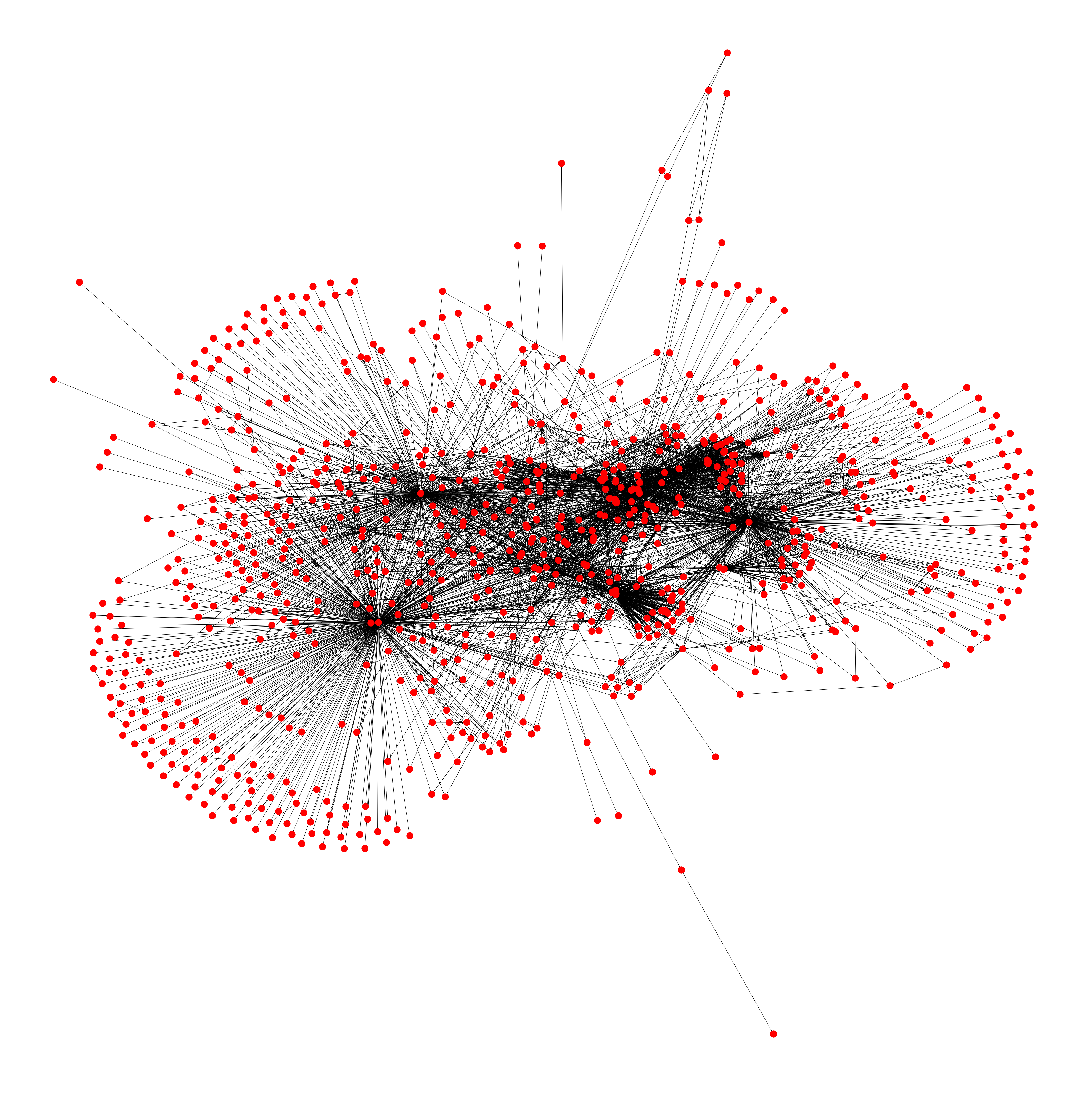}
    \end{center} 
    \caption{An example of social interaction graph generated from our data. 
    The number  of nodes equals the number of unique people participating in issue conversation.
    The existence and width  of each edge represent the frequency of conversation between pairs of developers. Hero programmers are those nodes that have very high node degree (i.e. who have participated in a lot of unique conversations). Note that,   these hero
    programmers are few in number.}
    \label{fig:Socialgraph}
\end{figure}

\subsection{Definitions}
Before beginning, we make some definitional points. When we say 1000+ projects, that is shorthand for the following. Our results used the intersection of two graphs of  {\em code interaction graph} (of who writes what and whose code) with {\em social interaction} graph (who discusses what issue and with whom) from 1037 projects.  Secondly, by code interaction graphs and  social interaction graphs, we mean the following. Each graph has   nodes and edges $\{N,E\}$. For code interaction graphs:
\bi
    \item Individual developers have their own  node $N_c$;
    \item The edge $E_c$ connects two nodes and indicates if ever one  developer has changed another developer's code. $W_c$ denotes how much one developer has changed another developer's code.
\ei
For social interaction graphs like \fig{Socialgraph}:
\bi
    \item A node $N_s$ is created for each  individual who has created or commented on an issue.
    \item An edge $E_s$ indicates communication between two individuals (as recorded in the issue tracking system.) If this happens $N$ times then the weight $W_s=N$.
\ei
Thirdly, we have defined heroes based on code contribution and communication. From the ``Code interaction graph'', the developers who contribute more than 80\% are hero contributors, and in the ``Social interaction graph'', the developers who are making 80\% of the communication are hero communicators. Both are ``hero developers'' for us. As we show in \S\ref{sec:Results}, both these definitions of heroes are insightful since more can be predicted about a project using {\em both} definitions that it either is applied separately.

\section{Background And Prior Work }
\label{sec:Background And Prior Work}

\subsection{Herbsleb Hypothesis (and Analogs)}
  At his ICSE'14 keynote, James Hersleb defined coding to be a socio-technical process where code and humans interact. According to the Hersleb hypothesis~\cite{Herbsleb:2014}, the following anti-pattern is a strong predictor for defects:
\bi
    \item If two code sections communicate...
    \item But the programmers of those two sections do not...
    \item Then that code section is more likely to be buggy.
\ei
To say that another way:

\begin{quote}
{\em Coding is a social process and better code arises from better social interactions. }
\end{quote}

Communication plays an important role in software development projects and the quality of communication has been found as determinant of project success~\cite{curtis1988field, kraut1995coordination, datta2021understanding}. The dynamic nature of open source software development makes the communication process between developers volatile~\cite{cataldo2007coordination}, consequently affecting a teams ability to effectively communicate and coordinate. These difficulties are more prominent in platforms like Github where we have distributed teams as developers work in geographically remote environments. The importance of communication in successful project completion is also well documented~\cite{griffin1992patterns, grinter1999geography, de2004good}. Many other researchers offer conclusions analogous to the Herbsleb hypothesis. Developer communication/interaction is often cited as one of the most important factors for a successful software development \cite{Agile_software_development,Kraut:1995:CSD:203330.203345,1205177, vale2020relation}. Many researchers have shown that successful communication between developers and adequate knowledge about the system plays a key role in successful software development \cite{TESCH2009657,Girba,841783}. As reported as early as 1975 in Brooks et al.  text ``The Mythical Man Month''~\cite{brooks1995mythical}, communication failure can lead to coordination problems, lack of system knowledge in the projects as discussed by Brooks et al. in the Mythical Man-Month. 

The usual response to the above argument is to improve communication by ``smoothing it out'', i.e. by deprecating heroes since, it is argued,  that encourages more communication across an entire project~\cite{bier2011online,boehm2006view,hislop2002integrating,morcovcomplex,wood2005multiview}.  The premise of ``smoothing it out'' is that heroes are bad and should be deprecated. This paper tries to verify whether or not this premise   holds true for open source GitHub projects or not.

\subsection{Heroism in Software Development}

Heroism in software development is a widely studied topic. Various researchers have found the presence of heroes in software projects. For example, In  1975  Brooks~\cite{Brooks:1975} proposed basing programming teams around a small number of ``chief programmers'' (which we would call ``heroes'')  who are supported by a large number of support staff (Brooks's analogy was the operating theater where one surgeon is supported by one or two anesthetists, several nurses, clerical staff, etc).  Also, the Agile Alliance~\cite{cockburn2006agile} and Bach et al.~\cite{bach1995enough} believed that heroes are the core ingredients in successful software projects saying ``... the central issue is the human processor - the hero  who steps up and solves the problems that lie between a need expressed and a need fulfilled.''  In 2002, Mockus et al.~\cite{mockus2002two} analyzed Apache and Mozilla projects to show the presence of heroes in those projects and reported,  their positive influence on  projects.  Also in 2002, Koch et al.~\cite{koch2002effort} studied the  GNOME project and showed the presence of heroes throughout the project history. They conjectured (without proof) that  the small number of hero developers may allow easy communication and collaboration. Interestingly, they also showed there is no relation between a developer's time in the project and being a hero developer.  In 2005, Krishnamurthy~\cite{KrishnamurthyS} studied 100 open source projects to find that a few individuals are responsible for the main contribution of the project in most of the cases.  In 2006 and 2009, Robles et al.~\cite{robles2009evolution,robles2006contributor} explored in their research the presence and evolution of heroes in open source software community.
In 2013, Peterson analyzed the software development process on GitHub and found out a pattern that most development is done by a small group of developers \cite{Peterson}. He stated that for most of the GitHub projects, 95-100\% of commits come from very few developers.  Also, as mentioned in the introduction,  In 2018, Agarwal et al. \cite{Agrawal_2018} stated that hero projects are very common. As software projects grow in size, nearly all projects become hero projects.

\renewcommand{\xmark}{}

\begin{figure*}
\includegraphics[width=\linewidth]{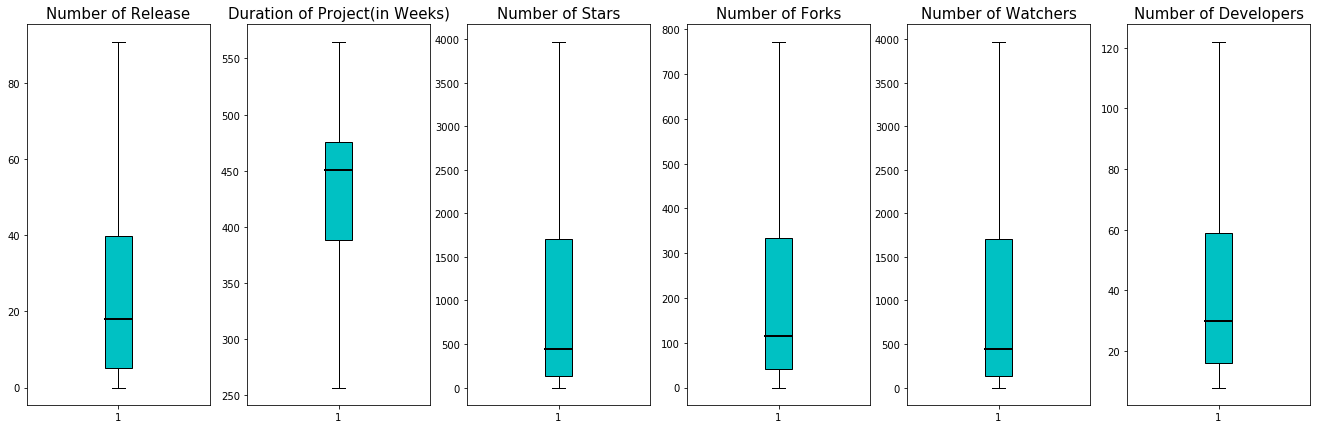}
\caption{Distribution of projects depending on Number of Releases, Duration of Project, Number of Stars, Forks. Watchers and Developers.  Box plots show the min to max range. Central boxes show the 25th, 50th, 75th percentiles.}
\label{fig:data}
\end{figure*}

Most prior researchers deprecate heroism in software projects arguing that:
\bi
    \item Having most of the work is dependent on a small number of heroes can become a bottleneck that slows down project development ~\cite{bier2011online,morcovcomplex,hislop2002integrating,boehm2006view,wood2005multiview}.
    \item In  the case of hero projects, there is less collaboration between team members since there are few active team members. So, it is argued, heroes are  negatively affecting the collaboration which is essential to the process of software development (and, more importantly, open source software development)\cite{1008000,4221622}. 
\ei

This second point is   problematic since, in the literature, studies that analyze  distributed software development on social coding platforms like GitHub and Bitbucket~\cite{dias2016does,cosentino2017systematic} remark on how social collaborations can reduce the cost and efforts of software development without degrading the quality of software. Distributed coding effort is beneficial for agile community-based programming practices which can in turn have higher customer satisfaction, lower defect rates, and faster development times ~\cite{moniruzzaman2013comparative,rastogi2017empirical}. Customer satisfaction, it is argued,  is increased when faster development leads to:

\bi
    \item Increasing the number of issues/bugs/enhancements being resolved~\cite{mockus2002two,jarczyk2014github,bissyande2013got,athanasiou2014test,gupta2014process,reyes2017analyzing}.
    \item Lowering the issues/bugs/enhancements resolution times \cite{jarczyk2014github, datta2021understanding}.
\ei

Even more specifically, as to issues related to heroes, Bier et al. warn when a project becomes complicated, it is always better to have a community of experts rather than having very few hero developers ~\cite{bier2011online}. Willams et al. have shown that hero programmers are often responsible for poorly documented software systems as they remain more involved in coding rather than writing code related documents ~\cite{hislop2002integrating}.  Also, Wood et al.~\cite{wood2005multiview} caution that heroes are often code-focused but software development needs workers acting as more than just coders (testers, documentation authors, user-experience analysts).

Our summary of the above is as follows: with only isolated exceptions, most of the literature deprecates heroes. Yet as discussed in the introduction, many studies indicate that heroic projects are quite common. This mismatch between established theory and a widely observed empirical effect prompted a re-examination of previous beliefs and the analysis discussed in this paper.

\section{Methodology}
\label{sec:Methodology}

Now to understand how communication between developers affects code quality, we need to perform different steps to - 

\bi
    \item Collect data about which developers coded which part of a project.
    \item Collect data about how each developer's code interacted with other developers.
    \item Collect data about how developers communicated with each other socially.
    \item Identify developers who contributes more and who contributes less both in-terms of code and communication.
    \item Collect and identify presence of bugs in a code base and identify who is responsible for the bug. 
\ei
The following section describes how we performed each steps in details.

\subsection{Data Collection}
\label{sec:Data Collections}

\fig{data} summarizes the Github data we used for this study.  To understand this figure, we offer the following definitions:

\bi
    \item {\em Release:} (based on Git tags)  marks a specific point in the repository's history. The number of releases defines different versions published, which signifies a considerable amount of changes done between each version.
    \item {\em Duration:} length of a project from its inception to the current date or project archive date. It signifies how long a project has been running and in the active development phase.
    \item {\em Stars:} signifies the number of people liking a project or use them as bookmarks so they can follow what's going on with the project later.
    \item {\em Forks:} A fork is a copy of a repository. Forking a repository allows users to freely experiment with changes without affecting the original project. This signifies how people are interested in the repository and actively thinking of modification of the original version.
    \item {\em Watcher:} Watchers are GitHub users who have asked to be notified of activity in a repository, but have not become collaborators. This is a representative of people actively monitoring projects, because of possible interest or dependency.
    \item {\em Developer:} Developers are the contributors to a project, who work on some code, and submit the code using commits to the codebase. The number of developers signifies the interest of developers in actively participating in the project and volume of the work.
\ei

\begin{figure}[!t]
\centering
{\scriptsize \renewcommand{\baselinestretch}{0.7}
\begin{tabular}{rrl}
    \textbf{Language} & \textbf{Projects} & \\
    Shell & 416 &\rule{86.17pt}{8pt} \\
    JavaScript & 396 &\rule{77.83pt}{8pt} \\
    HTML & 344 &\rule{76.67pt}{8pt} \\
    CSS & 314 &\rule{65.5pt}{8pt} \\
    Python & 291 &\rule{51pt}{8pt} \\
    Makefile & 229 &\rule{47.17pt}{8pt} \\
    Ruby & 216 &\rule{44.17pt}{8pt} \\
    C & 167 &\rule{41.83pt}{8pt} \\
    Java & 150 &\rule{39.67pt}{8pt} \\
    PHP & 146 &\rule{36.22pt}{8pt} \\
    C++ & 126 &\rule{34pt}{8pt} \\
    Batchfile & 81  &\rule{30.67pt}{8pt} \\
    Perl & 67 &\rule{28pt}{8pt} \\
    Objective-C & 67 &\rule{28pt}{8pt} \\
    Dockerfile & 54 &\rule{26.17pt}{8pt} \\
    CMake & 48 &\rule{22pt}{8pt} \\
    M4 & 43 &\rule{21.83pt}{8pt} \\
    CoffeeScript & 38 &\rule{19.83pt}{8pt} \\
    Roff & 35 &\rule{16.33pt}{8pt} \\
    Roff & 35 &\rule{16.67pt}{8pt} \\
    C-Sharp & 34 &\rule{15.67pt}{8pt} \\
    Emacs Lisp & 30 &\rule{14.5pt}{8pt} \\
    Gherkin & 26 &\rule{12.5pt}{8pt} \\
    Perl 6 & 21 &\rule{10.33pt}{8pt} \\
    
\end{tabular}}
\caption{Distribution of projects depending on languages.
Many  projects use combinations of languages to achieve their results. Here,  we show the languages which are predominantly used in the selected projects.}
\label{fig:lang_projects}
\end{figure}
\fig{lang_projects} shows that the projects we chose for our experiment comprise different languages. Note that we did not use all the data in Github. GitHub has over  100 million repositories as of the time of this collection so we only use data from the ``GitHub showcase project'' list. Many of these projects contain very short development cycles; are used for personal use; and are  not be related to software development. Such projects may bias research findings.  To mitigate that, we filter out projects using the standard ``sanity checks'' recommended in the literature~\cite{perils,curating}:  
\bi
    \item {\textit{{Collaboration}}: refers to the  number of pull requests. This is indicative of how many other peripheral developers work on this project. We required all  projects to have at least one pull request.}
    \item {\textit{{Commits}}: The project must contain more than 20 commits.}
    \item {\textit{{Duration}}: The project must contain software development activity of at least 50 weeks.}
    \item {\textit{{Issues}}: The project must contain more than 10 issues.}
    \item {\textit{{Personal Purpose}}: The project must have at least eight contributors.}
    \item {\textit{{Software Development}}: The project must be a placeholder for software development source code.}
    \item {\textit{Project Documentation Followed}: The projects should follow proper documentation standards to log proper commit comment and issue events to allow commit issue linkage.}
    \item {\textit{Social network validation}:  The Social Network that is being built should have at least 8 connected nodes in both the communication and code interaction graph (this point is discussed further in \ref{sec:Personnel Metrics} and \ref{sec:Product Merics}).}
\ei


For each of the selected project, the study recreates all the committed files to identify code changes in each commit file and identifies developers using the GitHub API, then downloads issue comments and events for a particular project, and uses the {\tt git log} command to mine the git commits added to the project throughout the project lifetime. Using the information from each commit message, this study uses keyword based identifier~\cite{rosen2015commit,hindle2008large,vasilescu2015quality} to label commits as buggy commit or not by identifying commits which were used to fix some bugs in the code and then  identifies the last commit which introduced the bug. This commit is labeled as a \textit{buggy commit}.

\subsection{Information Extraction}
\label{sec:Experimental Setup}
Here we discuss the process of extracting the relevant features and information regarding communication and code quality from the Github information collected~\cite{maruping2019developer, ortu2018mining, aljemabi2018empirical}

\subsubsection{Code Interaction Extraction}
\label{sec:process metrics}

Recall that the  developer code interaction graph records who touched what and whose code, where a developer is defined as a person who has ever committed any code into the codebase. We create that graph as follows:

\bi
    \item Project   commits  were extracted from each branch in git history. 
    \item Commits are extracted from the git log and stored in a file system.
    \item To access the file changes in each commit we recreate the files that were modified in each commit by (a)~continuously moving the {\tt git head} chronologically on each branch. Changes were then   identified  using {\tt git diff} on two consecutive git commits.
    \item The graph is created by going through each commit and adding a node for the committer. Then we use {\tt git blame} on the lines changed to find previous commits following a similar process to the SZZ algorithm~\cite{williams2008szz}. We identify all the developers of the commits from {\tt git blame} and add them as a node as well.
    \item After the nodes are created, directed edges were drawn between the developer who changed the code, to whose code was changed. Those  edges were weighted by the change size between the developers. 
\ei

\subsubsection{Social Interaction Extraction}
\label{sec:Personnel Metrics}

\begin{figure}[t]
\centering
\includegraphics[width=\linewidth]{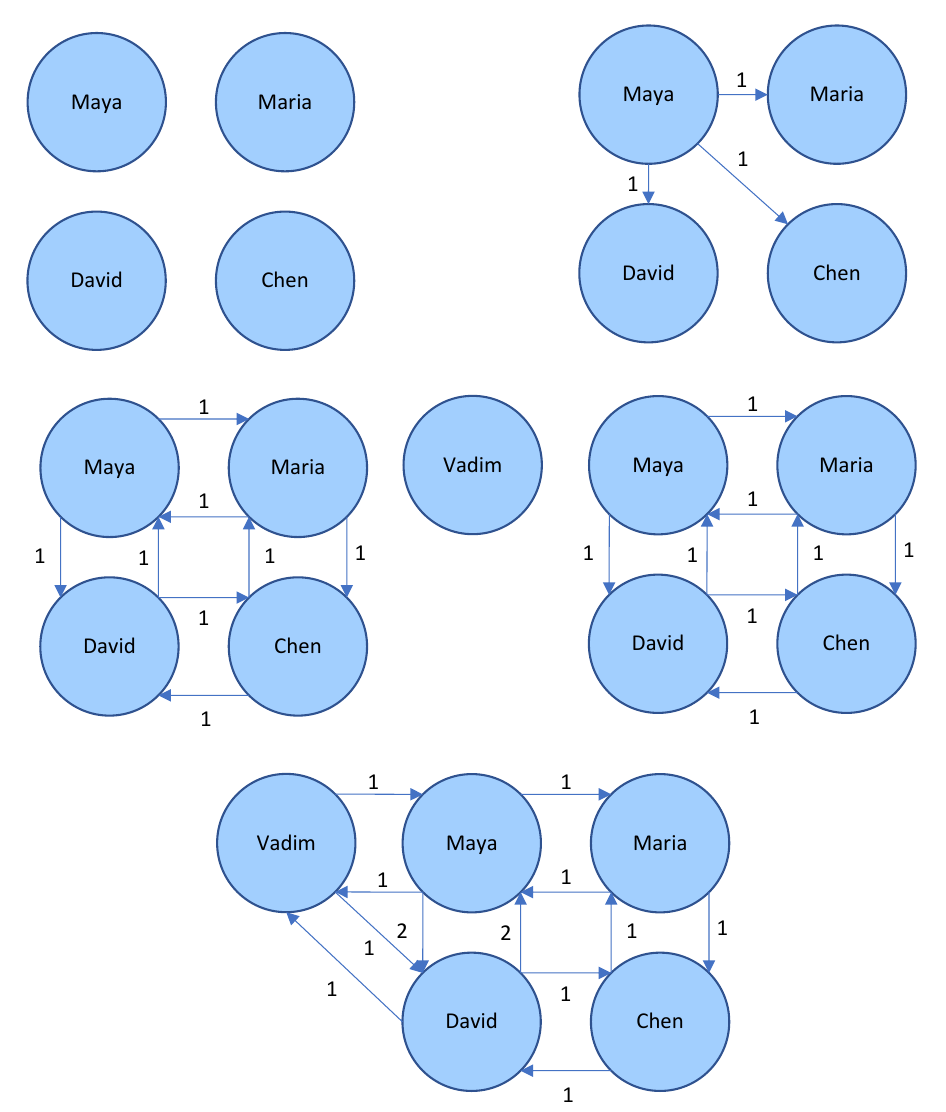}
\caption{\textbf{Example of creating a social interaction graph between four GitHub developers.}
\textbf{Step 1 (LHS):} Maya, Maria, Chen, and David are four developers in a GitHub project. \textbf{Step 2:} Maya creates one issue where Maria, Chen, and David comment. So, we join Maya-Maria, Maya-Chen, Maya-David with an edge of weight 1. \textbf{Step 3:} A new developer Vadim comes. \textbf{Step 4 (RHS):} Vadim creates one new issue where Maya and David comment. So, two new edges are introduced - (Maya-Vadim(1), David-Vadim(1)).  Now we iterate for each developer, so all of them become connected, and lastly,  the edge weight of Maya-David increases to 2.}
\label{fig:social_interaction_graph}
\end{figure}

Recall that the  developer social interaction graph records who talked to each other via issue comments. We create that graph as follows:

\bi
    \item A node is created for the person who has created the issue, then another set of nodes are created for each person who has commented on the issue. So essentially in the Social interaction graph, each node in the graph is any person (developer or non-developer) who has ever created an issue or commented on an issue.
    \item The nodes are connected by directed edges, which are created by (a) connecting from the person who has created the issue to all the persons who have commented on that issue and (b) creating edges from the person commenting on the issue to all other persons who have commented on the issue, including the person who has created the issue. 
    \item The edges are weighted by the number of comments by that person. 
    \item The weights are updated using the entire history of the projects, as per   Figure~\ref{fig:social_interaction_graph}.
\ei

\subsubsection{Defect Data Extraction}
\label{sec:Product Merics}

This study explores the effects of social and code communication to assess code quality, by measuring the percentage of buggy commits introduced by developers (hero and non-hero developers),  but to do so we do need to identify the commits that introduced the bug in the code from the historic project data. This is   a  challenging task since there is no direct way to find the commits or the person who is responsible for the bug/issue introduction. Hence, our scripts proceed as follows:

\bi
    \item Starting with all the commits from {\tt git log}, we identify the commit messages.
    \item Next,  to use the commit messages for labeling, we apply natural language processing~\cite{hindle2008large,rosen2015commit} (to do  stemming, lemmatization, and lowercase conversion to normalize the commit messages).
    \item Then to identify commit messages which are representation of bug/issue fixing commits, a list of words and phrases is  extracted from previous studies of 1000+ projects (Open Source and Enterprise). The system checked for these words and phrases in the commit messages and if found, it marks these as commits which fixed some bugs.
    \item To perform a sanity check,  5\% of the commits were manually verified by 7 graduate students using random sampling from different projects. Disagreements between manual labeling and keyword based labeling were further verified and keywords were added or removed to improve performance.
    \item These labeled commits were then processed to extract the file changes as the process mentioned in section~\ref{sec:process metrics}.
    \item Finally,  {\tt git blame} is used to go back in the git history to identify a responsible commit where each line was created or changed last time.
\ei

By this process, commits that were responsible for introduction of the bugs in the system/project can be found. We label these commits as ``buggy commits'' and label the author of the commit as the ``person responsible'' for introducing the bug.

\subsubsection{Final Feature Extraction}
\label{sec:Finding Relation}

To assess the prevalence of heroes in the software projects,  we joined all the metrics shown above. Specifically, using both social and code interaction graph, we calculated the node degree (number of edges touching a vertex) of the graphs (and note that vertices with a higher degree represent more communication or interaction). 

For the sake of completeness, we varied our threshold of ``hero'' to someone belonging to  80\%, 85\%, 90\%, and 95\% of the communication.  In our studies, top contributors (or heroes) and non-heroes
 were defined as : 
 
\begin{equation}
\label{eq:node-Degree}
    \mbox{ Node Degree of } N_i = D(N_i)= \sum_{j=1}^{n} a_{ij}
\end{equation}
\begin{equation}
\label{eq:hero}
    \mbox{  Hero} = Rank\left(D(N_i)\right) > \frac{P}{100}*(N + 1)
\end{equation}
\begin{equation}
\label{eq:non-hero}
    \mbox{  Non-Hero} = Rank\left(D(N_i)\right) < \frac{P}{100}*(N + 1)
\end{equation}
where:

\begin{table}[h]
\centering
\begin{tabular}{|l|l|}
\hline
N & Number of Developers \\ \hline
P & 80, 85, 90, and 95 Percentile \\ \hline
Rank() & \begin{tabular}[c]{@{}l@{}}The percentile rank of a score is the percentage \\ of scores in its frequency distribution that is\\ equal to or lower than it.\end{tabular} \\ \hline
a & \begin{tabular}[c]{@{}l@{}}Adjacency matrix for the graph \\ where $a_{ij}>0$  denotes a connection\end{tabular} \\ \hline
\end{tabular}
\end{table}

Using these data and by applying the hero definition from formula (\ref{eq:hero}) and (\ref{eq:non-hero}) (look at the top 20\%, 15\%, 10\%, and 5\%), we can find the developers who are responsible for 80\%, 85\%, 90\%, and 95\% of the work. We use this to categorize the developers into 2 groups:
\bi
    \item The \textit{hero developers}; i.e. the core group of the developers of a certain project who make regular changes in the codebase. In this study, this is represented by the developers whose node degree is above 80, 85, 90, and 95th percentile of the node degree (developers' communication and code interaction of the system graph).
    \item The \textit{non-hero developers} are all other developers; i.e.  developers associated with  nodes with a degree below the respective threshold percentile.
\ei

\begin{figure}[!b]
\includegraphics[width=\linewidth]{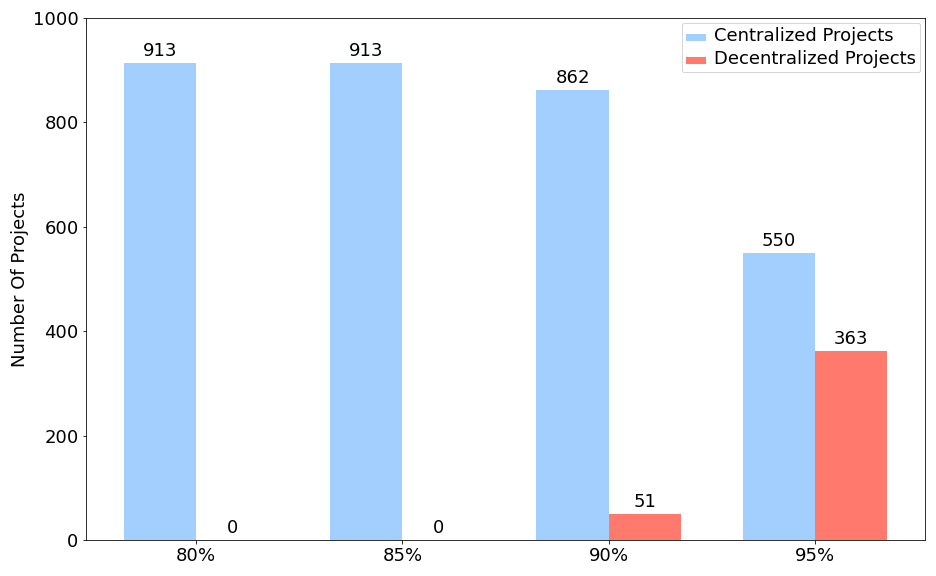}
\caption{RQ1 Result: Distribution of hero projects vs non-hero projects based on hero threshold being 80\%, 85\%, 90\%, and 95\% respectively. Here threshold being 80\% means in that project 80\% of the code is done by less than 20\% of developers}
\label{fig:RQ1_following}
\end{figure}
Following this for each selected project, we merge the data collected in section~\ref{sec:Product Merics} and section~\ref{sec:process metrics} to find each developer's code contribution according to the code interaction graph. A similar process is followed for  section~\ref{sec:Personnel Metrics} and section~\ref{sec:process metrics} in the social interaction graph. Using the above mentioned data, we can validate the code and social contribution of each developer along with their bug introduction percentage. This information will help us to answer the research questions asked in this study.

\section{Results}
\label{sec:Results}
Our results are structured around three research questions:
\begin{description}
    \item[{\bf RQ1:}] How common are projects with highly centralized social and code communication structure (i.e., hero projects)?
    \item[{\bf RQ2:}] What impact  does high social and code communication have  on  code quality?
    \item[{\bf RQ3:}] Do the results support Herbsleb Hypothesis?
\end{description}

\subsection*{\textbf{RQ1: \textit{How common are projects with highly centralized social and code communication structure (i.e., hero projects)?}}}

Figure~\ref{fig:RQ1_following} shows the result for the number of projects where majority of social and code communication is done by only a few developers. We mark these projects as  as hero and non-hero based on a varying  threshold of $x$ percentage of work (both social and code communication) done by $y$ percentage of developers. Here we vary the $x$ percentage (that is the work done) from 80 to 95\%, while we vary $y$ (i.e., percentage of developers) from 20 to 5\%. The clear conclusion from this figure is that the phenomenon  that we have defined as ``hero`` is very common. In fact, the phenomenon may be more pronounced than previously reported. Even when we require heroes to be  involved in 95\% of the communication  (which is a large amount), we find that majority of the projects studied here exhibits a very centralized social and code communication structure.

\begin{result}
Here we see a overwhelming presence of  projects with highly centralized social and code communication structure in open source software community. That is, in the usual case, there are very few people in each project responsible for majority of the work.
\end{result}

\subsection*{\textbf{RQ2: \textit{What impact  does high social and code communication have  on  code quality?}}}

RQ2 explores the effect of high social and code communication (heroism) have on code quality. In this study, we created the developer social interaction graph and developer code interaction graph, then identified the developer responsible for introducing those bugs into the codebase. Then we find the percentage of buggy commits introduced by those developers by checking (a)~the number of buggy commits introduced by those developers and (b)~their number of total commits.

\begin{figure}
\begin{center}
\includegraphics[width=\linewidth]{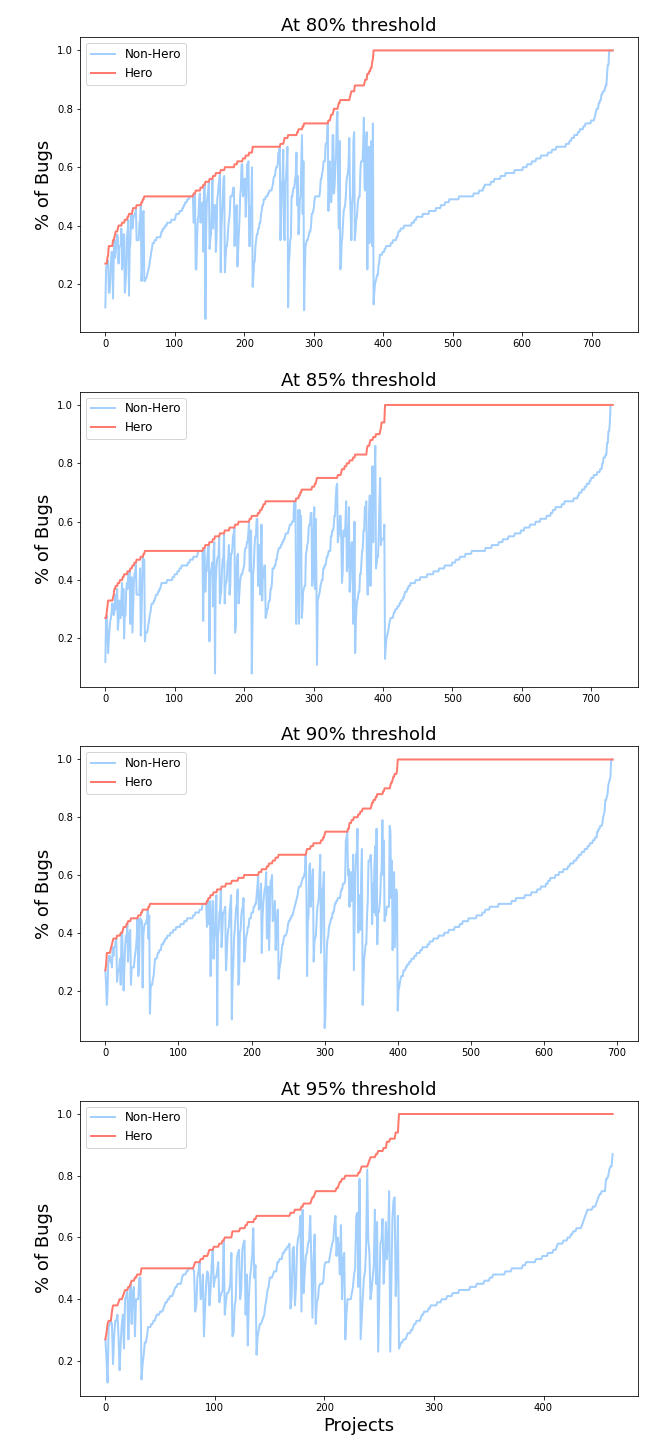}
\end{center}
\caption{Code interaction  graph results for
RQ2: Percentage of bugs introduced by hero and non-hero developers from developer code interaction perspective in Hero Projects.}
\label{fig:RQ2_code}
\end{figure}

\begin{figure}\begin{center}
\includegraphics[width=1.07\linewidth]{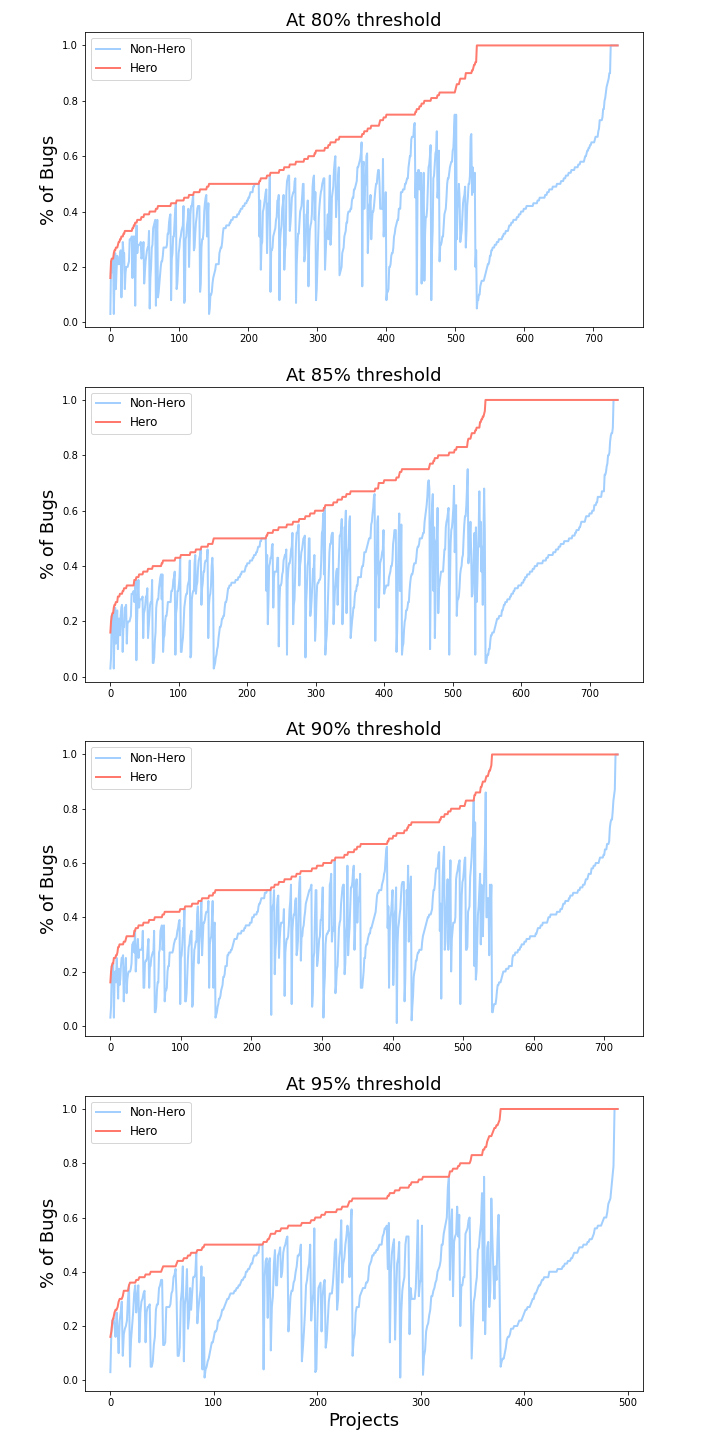}
\end{center}
\caption{Social interaction  graph results for
RQ2: Percentage of bugs introduced by hero and non-hero developers from developer social interaction perspective in Hero Projects.}
\label{fig:RQ2_social}
\end{figure}

Fig~\ref{fig:RQ2_code} and Fig~\ref{fig:RQ2_social}
shows the comparison between the performance developers who had made high social and code communication (marked as hero) vs who had made little social and code communication (marked as non-hero) to the project. In those figures:
\bi
\item
The x-axis is different projects used in this study.
\item
The y-axis represents the median of the bug introduction percentage for all hero and non-hero developers for each project respectively.
\ei


Here projects are sorted by the number of non-hero developers. In those charts, we note that:
\bi
\item
There exists a large number of non-heroes who always produce buggy commits, 100\% of the time (evidence: the flat right-hand-side regions of the non-hero plots in both figures). That population size of ``always buggy'' is around a third in 
Fig~\ref{fig:RQ2_code} and a fourth in Fig~\ref{fig:RQ2_social}.
\item
To say the least, heroes nearly always have fewer buggy commits than non-heroes. The 25th, 50th, 75th percentiles for both groups are
shown in Table~\ref{tbl:team_size}. This table clearly shows why heroes are so  prevalent, they generate commits that are dramatically
less buggy than non-heroes, regardless of the size of the project.
\ei

\begin{table}[]
\caption{The table summarizes of Fig~\ref{fig:RQ2_code}, Fig~\ref{fig:RQ2_social} and stratifies the data according to 25th, 50th, and 75th percentile of the developers.}
\label{tbl:team_size}
\small
\begin{center}
\begin{tabular}{|c|c|c|c|c|}
\hline
\rowcolor[HTML]{EFEFEF} 
\textbf{}                                     & \textbf{}         & \multicolumn{3}{c|}{\cellcolor[HTML]{EFEFEF}\textbf{Percentile}} \\ \hline
\rowcolor[HTML]{EFEFEF} 
\textbf{Metric}                               & \textbf{Group}    & \textbf{25th}        & \textbf{50th}       & \textbf{75th}       \\ \hline
                                              & \textbf{Hero}     & 0.52                 & 0.58                & 0.53                \\ \cline{2-5} 
                                              & \textbf{Non-Hero} & 0.67                  & 0.75                 & 1.0                 \\ \cline{2-5} 
\multirow{-3}{*}{\textbf{Code Interaction}}   & \textbf{Ratio}    & 1.3                  & 1.3                 & 1.9                 \\ \hline
                                              & \textbf{Hero}     & 0.44                  & 0.5                 & 0.5                 \\ \cline{2-5} 
                                              & \textbf{Non-Hero} & 0.67                 & 0.75                & 0.67                \\ \cline{2-5} 
\multirow{-3}{*}{\textbf{Social Interaction}} & \textbf{Ratio}    & 1.5                  & 1.5                 & 1.3                 \\ \hline
\end{tabular}
\end{center}
\end{table}

The other thing to note from 
Fig~\ref{fig:RQ2_code} and Fig~\ref{fig:RQ2_social}
is that they are nearly identical.
That is, no matter how we define ``hero'',
we reach the same conclusions.
Hence we say -  

\begin{result}
In modern software projects,
 reflecting on
  who writes most of the code is just as insightful
  as reflecting on  who participates in most
of the discussion about the code.
\end{result}

\subsection*{\textbf{RQ3: \textit{Do the results support Herbsleb Hypothesis?}}}

In this research question, we explored the Herbsleb hypothesis~\cite{Herbsleb:2014} from section~\ref{sec:Background And Prior Work}; i.e. does lack of communication between developers predict for bugs in the code? To do that for 1,037 projects, we discretized the developers into 3 groups (i.e. High, Medium, and Low) based on their code contribution (Code Node Degree) and social communication frequency (Social Node Degree). In Table~\ref{fig:Herbsleb_stats}, group HH (High, High) represents the developers who have high code contribution and social communication frequency and the value in the cell is median bug introduction percentage, while group LL (Low, Low) represents the developers who have low code contribution and low social communication frequency. 


Table~\ref{fig:Herbsleb_stats} shows the result of a statistical test performed on the 9 different groups representing different frequency and volume of communication.
The ``rank'' column of that table shows a statistical analysis where one row has a higher rank than the previous one only if the two rows are
\bi
\item
 Statistically {\em significant different.}
For this test, we used the Scott-Knot method recommended by Angelis and Mittas at TSE'13~\cite{mittas13}.
\item
And that difference is 
{\em not a small effect}. 
For this effect size test, we used the A12  test recommended by Angelis and Briand at ICSE'11~\cite{arcuri11} 
\ei
These statistical tests were selected
since they were non-parametric; i.e. they do not
assume normal distributions. 

To summarize the effect reported in Table~\ref{fig:Herbsleb_stats},
we need to look at the difference between the highest and
lowest ranks:
\bi
\item
In the groups with the highest defects and highest rank,
the social and coding groups are both low.
That is non-hero-ness (for both code and social interaction)
is associated with the worst quality code.
\item In the groups with the lowest defects and lowest rank,
the social group is always high while the code groups
are either low or medium. That is, extensive
social interaction even with low to medium code interaction
is associated with the best quality code.
\ei
From the above, we can say that:\
\bi
\item These results support the Herbsleb hypothesis, which suggests communication between developers is an important factor and lesser social communication can lead to more bugs.
\item Also, prior definitions of ``hero'' based just on code interaction need now to be augmented. As shown in Table~\ref{fig:Herbsleb_stats}, social hero-ness is much more predictive for bugs than merely reflecting. on code hero-ness.
\ei

This finding leads to  the following
conjecture (to be explored
in future work):
the best way to reduce communication overhead and to decrease defects is to {\em centralize the communicators}. In our data, commits with lower defects come from the small number of hero developers who have learned how to talk to more people. Hence, we would encourage more research into better methods for rapid, high-volume, communication in a one-to-many setting (where the ``one'' is the hero and the ``many'' are everyone else). In summary, we can say 

\begin{result}
The Herbsleb hypothesis holds true for open source software projects. We see that when developer's codes interact with each other, if there is significant social interaction (social communication) between them as well they tends to introduce much less bugs into the code base than when there is not. And more research should be performed to find better methods for rapid, high-volume, communication in a one-to-many setting.
\end{result}



\begin{table}[!t]
\centering
\caption{This figure shows the result of statistical significance test and an effect size test on 9 different groups used to study the Herbsleb hypothesis. In this figure the ``group'' column represents the 9 different groups in this research question, where the first character represents the code node degree, while the later is social node degree.
}
{\small
{\small \begin{tabular}{lcrrp{2.5cm}}
\arrayrulecolor{darkgray}
\rowcolor{gray!30}  rank & \begin{tabular}[c]{@{}c@{}}group\\ (code,social)\end{tabular} & median & IQR & \\
 
  1 &      L,H &    30 &  29 & \quart{18}{30}{29}{17} \\
  1 &      M,H &    38 &  28 & \quart{26}{38}{28}{16} \\ \hline
  2 &      L,M &    38 &  31 & \quart{25}{38}{31}{18} \\
  2 &      H,H &    42 &  18 & \quart{33}{42}{18}{9} \\
  2 &      M,M &    46 &  21 & \quart{35}{46}{21}{10} \\
  2 &      H,M &    46 &  21 & \quart{33}{46}{21}{8} \\
  2 &      H,L &    48 &  30 & \quart{34}{48}{30}{16} \\ \hline
  3 &      M,L &    52 &  29 & \quart{38}{52}{29}{15} \\ \hline
  4 &      L,L &    67 &  45 & \quart{35}{67}{45}{23} \\
\end{tabular}}
}

\label{fig:Herbsleb_stats}
\end{table}

\section{Threats to Validity}
\label{sec:Threats to Validity}

\subsection{Sampling Bias} Our conclusions are based on 1000+ open source GitHub projects that started this analysis.  It is possible that   different initial projects would have lead to different conclusions. That said, our initial sample is very large so we have some confidence that this sample represents an interesting range of projects.  
            
\subsection{Evaluation Bias} In  RQ1, RQ2, and RQ3, we said that projects with highly centralized social and code communication structure very common, that is to say hero projects are common in open source community. Developers who are responsible for majority of the work (both social and code) are responsible for far less bug introduction than non-hero developers. It is possible that using other metrics\footnote{E.g. do  heroes reduce productivity by becoming bottleneck?} then there may well be a difference in these different kinds of projects. But measuring people resources only by how fast releases are done or issues are fixed may not be a good indicator of having heroes in a team. This is a matter that needs to be explored in future research. 
     
Another evaluation bias as we report cumulative statistics of lift curves where other papers reported precision and recall. The research in this field is not mature enough yet for us to say that the best way to represent results is one way versus another. Here we decided to use lift curves since, if we used precision and recall, we had to repeat that analysis at multiple points of the lift curve. We find our current lift curves are a succinct way to represent our results.

\subsection{Construct Validity} At various places in this report, we made engineering decisions about (e.g.) team size; and (e.g.) what constitutes a successful project. While those decisions were made using advice from the literature (e.g.~\cite{gautam2017empirical}), we acknowledge that other constructs might lead to different conclusions. 
     
That said,  while we cannot prove that all of our constructs are  in any sense ``optimal'',  the results of Table~\ref{fig:Herbsleb_stats} suggest that our new definition of social hero-ness can be more informative than constructs used previously in the literature (that defined ``hero'' only in terms of code interaction).  
         
Another issue about our construct validity is that we have relied on a natural language processor to analyze commit messages to mark them as buggy commits. These commit messages are created by the developers, and may or may not contain a proper indication of if they were used to fix some bugs. There is also a possibility that the team of that project might be using different syntax to enter in commit messages.

Yet another threat to construct validity is that we did not consider the different roles of the developers. We had trouble extracting that information from our data source, we found that people have multiple roles particularly our heroes who would often step in and assist in multiple activities.  Nevertheless the exploration of different roles would be an interesting study.

\subsection{External Validity} 
Previously Agrawal et al. were able to comment on the effects of heroes in  open and closed source projects. That research group was fortunate enough to work on-site at a large open source software company. We were not as fortunate as them. We, therefore, acknowledge our findings (from open source projects) may not be the same for closed source projects.  


Similarly, we have used GitHub issues and comments to create the communication graph, it is possible that the communication was not made using these online forums and was done with some other medium. To reduce the impact of this problem, we  did take precautionary steps to (e.g.,) include various tag identifiers of bug fixing commits, did some spot check on projects regarding communication, etc.

Our conclusion shows that almost all (when experimenting with 80\%, 85\%, 90\% threshold) of our sample projects are hero dominated.  In case of large size public GitHub projects, there are official administrators and maintainers who are responsible for issue labeling or assigning. So, they frequently comment on all of the issues but though they are not active developers. These people should not be considered hero developers. Finding these people needs manual inspection which is not possible for 1000+ projects. We decide to put it as a limitation of our study as we deal with a huge number of projects.

We do not isolate hero projects and non-hero projects and look into them separately because there are very few non-hero projects and also there are a lot of developers who work on a large number of projects (some of them are hero projects and some of them are not). 


\section{Conclusion}
\label{sec:Conclusion}

In this study we investigated the impact of social and code communication on software code quality, expressed through their impact on percentage of defects introduced by developers who had made significant social and code communication (hero developers) vs  who had made insignificant social and code communication (non-hero developers). The established wisdom in literature expresses that communication between developers is a key aspect of a successful project and also the established wisdom in the literature is to depreciate ``heroes'', i.e., a small percentage of the staff who are responsible for most of the progress on a project. As James Herbsleb~\cite{Herbsleb:2014}, in his ICSE’14 keynote said - ``If two code sections communicate but the programmers of those two sections do not, then that code section is more likely to be buggy.''. Then the question arises, what is the effect on high or low  social and code communication have on code quality, that is  to say what effect of high social and code communication (i.e, heroism) have on code quality?  Based on our study of 1000+ open source GitHub projects,
we assert:
\bi
    \item Overwhelmingly, most projects have highly centralized social and code communication structure. That basically means, in majority of the projects there exists only a few developers who are responsible for majority of social and code communication. 
    \item Developers who are responsible for majority of the social and code communication are  far less likely to introduce bugs into the codebase than their counterparts. Thus showing developers who are responsible for majority of the social and code communication (that is hero developers) are beneficial towards the project quality and both social and  code communication between developers have significant effect on code quality. 
    \item Our experiments empirically supports the Herbsleb Hypothesis, showing - ``If two code sections communicate but the programmers of those two sections do not, then that code section is more likely to be buggy.''. 
\ei

One strange feature of our results is that what is old is now new. Our results (that heroes are important) echo a decade-old concept. In 1975, Fred Brooks wrote of  ``surgical teams'' and the ``chief programmer'' ~\cite{brooks1974mythical}. He argued that:

\bi
  \item Much as a surgical team during surgery is led by one surgeon performing the most critical work while directing the team to assist with less critical parts.
  \item Similarly, software projects should be led by one or a few ``chief programmer'' to  develop critical system components while the rest of a team provides what is needed at the right time.
\ei

Brooks conjecture that ``good'' programmers are generally much more as productive as mediocre ones. This can be seen in the results that programmers are much more productive (who are responsible for majority of the social and code communication) and less likely to introduce bugs into the codebase. These developers are born when they become so skilled at what they do, that  they assume a central position in a project.
In our view,  organizations need to acknowledge their dependency on such developers, perhaps altering their human resource policies and manage these people more efficiently by retaining them.

Also, our results suggest one other way to the way we develop new technologies or modern software projects. Specifically,  given the prominence and importance of social and code communication between developers, future work could usefully explore methods  to streamline the communication between developers that are critical for high quality software.

Finally, while this paper  has been about how. social and code communication affects code quality,  there might be a  more general point to be made here: {\em it is time to reflect more on long-held truisms in our field}. While Heroes (developers who are responsible for majority of the social and code communication) are widely deprecated in the literature, yet empirically they are quite beneficial. What other statements in the literature need to be reviewed and revised?

\section*{Acknowledgements}
This research was partially funded by  NSF Grant \#1908762.

\bibliographystyle{elsarticle-num}
\bibliography{Mybib}

\begin{thebibliography}{10}
\bibitem{bach1995enough}
J.~Bach, Enough about process: what we need are heroes, IEEE Software 12~(2).
\end{thebibliography}

\balance
\end{document}